\documentclass[fleqn,10pt]{wlscirep}
\title{(geo)graphs - Complex Networks as a shapefile of nodes and a shapefile of edges for different applications}

\author[1,2,*]{Leonardo B L Santos}
\author[3]{Aurelienne A S Jorge}
\author[4]{Marcio Rossato}
\author[2]{Jéssica D Santos}
\author[1]{Onofre A Candido}
\author[5]{Wilson Seron}
\author[6,7,8]{Charles N de Santana}

\affil[1]{Centro Nacional de Monitoramento e Alertas de Desastres Naturais (Cemaden), S\~ao Jos\'e dos Campos, SP, Brazil}
\affil[2]{Programa de  P\'os Gradua\c{c}\~ao em Computação Aplicada do Instituto Nacional de Pesquisas Espaciais (INPE), São José dos Campos, SP, Brasil}
\affil[3]{Instituto Nacional de Pesquisas Espaciais (INPE), Cachoeira Paulista, SP, Brasil}
\affil[4]{Via Vale Sistemas, Cachoeira Paulista, SP, Brasil}
\affil[5]{Universidade Federal de São Paulo, São José dos Campos, SP, Brasil}
\affil[6]{Instituto Nacional de Ci\^encia e Tecnologia em Estudos Interdisciplinares e Transdisciplinares em Ecologia e Evolu\c{c}\~ao (INCT-INTREE), Universidade Federal da Bahia, Salvador, BA, Brasil}
\affil[7]{Programa de P\'os Gradua\c{c}\~ao em Modelagem em Ci\^encias da Terra e do Ambiente, Universidade Estadual de Feira de Santana, Feira, BA, Brasil}
\affil[8]{Departamento de Computação Científica, DataSCOUT - Data Science and Scientific Computing, Salvador, BA, Brasil}

\affil[*]{leonardo.santos@cemaden.gov.br}

\keywords{spatial networks, geographical networks, urban mobility, meteorological networks'}

\begin{abstract}
Spatial dependency and spatial embedding are basic physical properties of many phenomena modeled by networks. The most indicated computational environment to deal with spatial information is to use Georeferenced Information System (GIS) and Geographical Database Management Systems (GDBMS). Several models have been proposed in this direction, however there is a gap in the literature in generic frameworks for working with Complex Networks in GIS/GDBMS environments. Here we introduce the concept of (geo)graphs: graphs in which the nodes have a known geographical location and the edges have spatial dependence. We present case studies and two open source softwares (GIS4GRAPH and GeoCNet) that indicate how to retrieve networks from GIS data and how to represent networks over GIS data by using (geo)graphs.
\end{abstract}
\begin{document}

\flushbottom
\maketitle
%
%
\thispagestyle{empty}


\section*{Introduction}
The advances in high computing processing and data storage in last decades together with the development of modern statistical physics theory facilitated the arise of Complex Networks approaches to model large scale systems in which it is possible to find relationships among its elements \cite{watts1998collective, barabasi1999emergence, strogatz2001exploring, albert2002statistical}. This concept has been applied to interdisciplinary studies ranging from Ecology\cite{sole2001complexity, pascual2006ecological}, Evolution\cite{wagner2001small, andrade2011detecting, wagner2011genotype}, Urban Mobility\cite{Barbosa:2017}, Epidemiology\cite{pastor2001epidemic}, Computer Sciences \cite{barabasi2000scale}, Climatic Sciences \cite{donges2009complex, de2009graph} among others.

Spatial dependency and spatial embedding are basic physical properties of many phenomena modeled by networks. The classical approach to deal with spatial information is to use Georeferenced Information System (GIS) and Geographical Database Management Systems (GDBMS) that easily combine different layers of georeferenced data into information retrieval. Several models have been proposed to describe spatial effects in networks \cite{hayashi2006review, barthelemy2011spatial}. However, there is a gap in the literature about generic frameworks for working with Complex Networks in GIS/GDBMS environments.


Here we introduce the concept of (geo)graphs: graphs in which the nodes have a known geographical location and the edges have spatial dependence - and (geo)graphs are geographical objects compatible with GIS/GDBMS environments. We also present two open source computational frameworks to facilitate the integration of the (geo)graphs in GIS/GDBMS environments: \emph{GIS4GRAPH} (G4G), a Web tool written in Python and JavaScript that allows the conversion of a given shapefile to a (geo)graph, for which the only requirements to use it is to have Internet access and a browser; and \emph{Geographical Complex Networks} (GeoCNet) a desktop application written in Python and C that allows the conversion of a graph to a (geo)graph shapefile. Both, G4G and GeoCNet, have been developed by some of the authors of this paper, and both source codes are free and can be accessed, respectively, in the repositories \url{https://github.com/aurelienne/gis4graph} and \url{https://github.com/jessicadominguess/geocnet}.


\section*{The G4G tool}

\subsection*{From GIS to graphs - the G4G sofware}

In terms of geospatial data, GIS4Graph is able to deal with both shapefiles and OpenStreetMap (OSM) files as input. When dealing with a shapefile, it must be a set of linestrings representing the network to be analyzed. Such data are then inserted into a database with geographic support - using PostgreSQL as the Database Management System and PostGIS as its spatial extension.  The connections identification between network segments is efficiently performed by an indexed spatial query based on a function that verifies intersections between geometric features. When it comes to an OSM file representing a street network, a PostgreSQL extension named pgRouting is employed. In both cases, the result is a connection list between nodes.

The first step before any calculation is to build a graph based on the analyzed network. It can be done by adding a node for each geographic feature and edges based on the connection list. The igraph library \cite{igraph2006} is used for both graph building and some default metrics calculation. Vertex degree, clustering coefficient, shortest paths and betweenness are some examples of metrics calculated by igraph and incorporated to G4G. G4G interface is built upon OpenLayers, a JavaScript framework to work with interactive maps and geographic elements, allowing applying visualization styles to features individually. The resultant dataset is displayed visually on a map or as a graph.

\subsection*{From graphs to GIS - the GeoCNet software}

In order to convert spatial networks to GIS environment we propose the following workflow:

\begin{enumerate}
\item To create a shapefile for the nodes using any GIS software. A \emph{point type shapefile} for the nodes must be created. The shapefile must have a mandatory column of type integer named \emph{id}, representing the id's of the nodes. All the characteristics of the polygons/points will be associated to their respective points as attributes, including the geographic locations of the nodes. The resulting nodes shapefile will be used as an \textbf{input} for the GeoCNet software.

\item To create an adjacency matrix (0s and 1s) representing the connections between these nodes. The matrix will be used as an \textbf{input} for the GeoCNet software as well.

\item Then, a \emph{line type shapefile} representing the edges of the network is given as an \textbf{output} of GeoCNet. The point-type-shapefile and the line-type-shapefile will have topological attributes of nodes and edges respectively.
\end{enumerate}


\section*{Some Applications}

\subsection*{Transportation}

Here we present how (geo)graphs can be used to identify potential main streets used by most of the transports routes in the city of Lorena/SP. The geodata representing the street network of the city of Lorena/SP were acquired through a request on OpenStreetMap Extended API by specifying the bounding box of the city. It delivers an XML response wrapped in an OSM element that includes basically the description of the ways (polylines that represent linear features such as roads) and their relationships (OSM, 2017). More precisely, each line segment between crossroads is a way, and the relationships between ways are indicated by ‘osm\_source’ and ‘osm\_target’ fields. For the proposed analysis, it is needed to represent every avenue or street as a single node.

\begin{figure}[!ht]
\centering
\includegraphics[width=\linewidth]{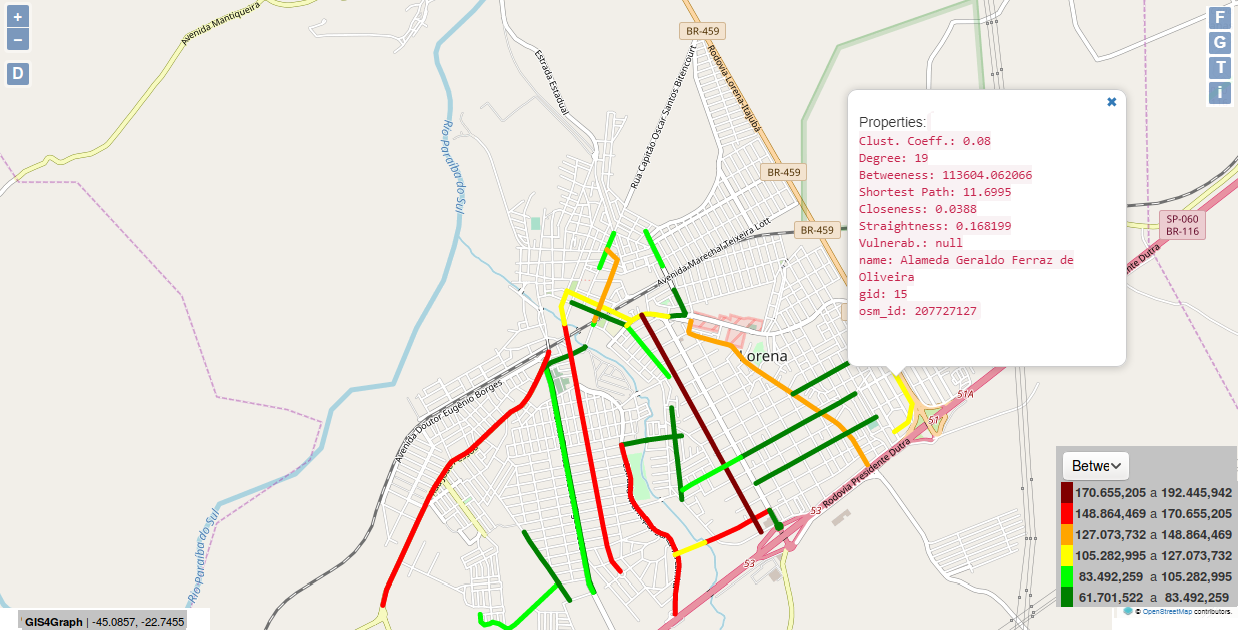}
\caption{Lorena city's street network on map visualization. A filter was applied to show only the streets with the highest betweenness values with colors employed according to the legend on the bottom right corner.}
\label{fig:bet_map}
\end{figure}

By using the G4G, we can extract the network from the GIS data and then represent the betweenness centrality index for the streets network on a map. A case study for the city of Lorena/SP is shown in Figure \ref{fig:bet_map}. The streets for which the betweenness centrality index are higher are the potential main streets used by most of the transport routes in the city.

\subsection*{Mobility - flow}
We studied the flow of people between each pair of areas (traffic zones) in a city on a typical day using an Origin-Destination survey for the metropolitan region of Rio de Janeiro/RJ \cite{TTC}.  

\begin{figure}[!ht]
\centering
\includegraphics[width=\linewidth]{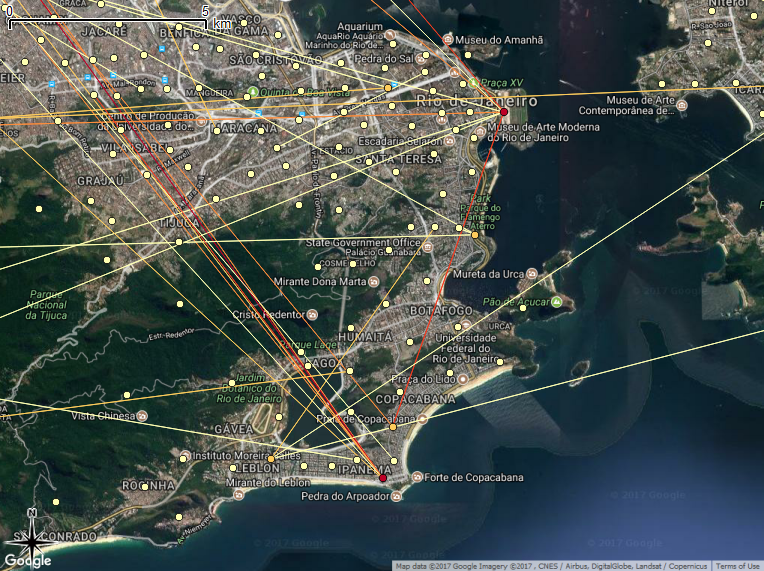}
\caption{Urban mobility network for the metropolitan region of Rio de Janeiro/Brazil. The color of each node is related to its topological degree (number of connections), and the color of each edge is related to its weight index (total number of people going from one to the other node of each pair of nodes). The threshold of flow to connect each pair of nodes was taken in a thousand people.}
\label{fig:mob}
\end{figure}

By using the GeoCNet, we can represent urban mobility networks on a map. A case study for the metropolitan region of Rio de Janeiro/Brazil is shown in Figure \ref{fig:mob}. Each node represents a traffic zone. Each pair of nodes is connected if the flow of people between the traffic zones represented by them is greater than a given threshold value. Nodes colors change according to the number of connections of the nodes. And edges colors change according to the total number of people going from one to the other node. In the figure it is shown that many people move between distant traffic zones (red edges between nodes distant from each other) especially if the destination/origin is a very connected traffic zone (a red node).

\subsection*{Weather radar - correlation}
Weather radar is one of the most important equipment for monitoring extreme meteorological events. For this brief case study, a dataset of meteorological time series were obtained by a weather radar (temporal resolution of 10 minutes)\footnote{sigma.cptec.inpe.br/radar/}.

\begin{figure}[!ht]
\centering
\includegraphics[width=\linewidth]{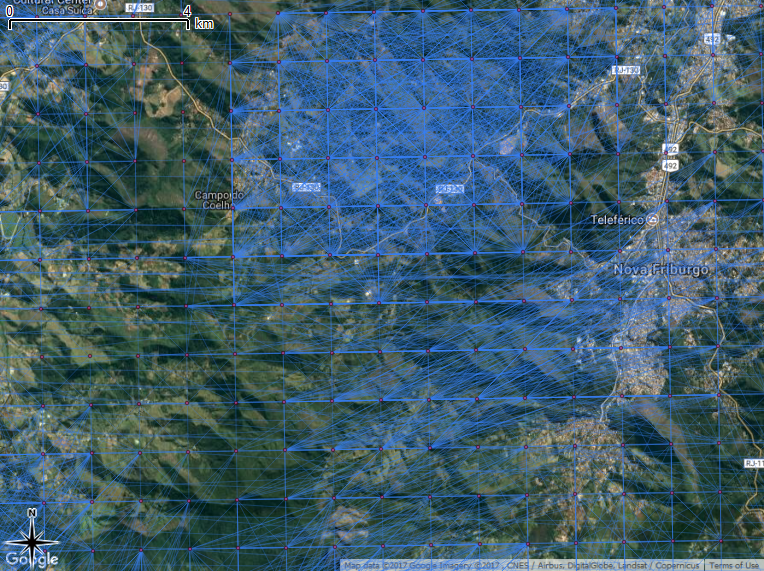}
\caption{Correlation network for the rainfall time series in the mountains region of Rio de Janeiro/Brazil. The satellite image of the region is also represented.}
\label{fig:rain}
\end{figure}

By using the GeoCNet, we can study rainfall temporal correlation networks. A case study for the mountainous region of Rio de Janeiro/Brazil, near the city Nova Friburgo/Brazil, is shown in Figure \ref{fig:rain}. Each node represents a grid point of a weather radar (as an interpolated field of rainfall). Each pair of nodes is connected if the Pearson correlation between the rainfall time series associated to each node is greater than a given threshold value. In the Figure, red dots represent the nodes and the blue lines represent the edges. So it is possible to visualize a community-like structure in the network, guided by the mountains of the region (see the satellite image).

\section*{Conclusions and Perspectives}

In this paper it was introduced the concept of (geo)graphs: graphs in which the nodes have a known geographical location and the edges have spatial dependence. We also presented two open source softwares GIS4GRAPH (G4G) and GeoCNet, two computational frameworks to facilitate the integration of the (geo)graphs in GIS softwares. Datasets about transportation (streets - OSM), urban mobility (flow of people) and meteorological (time series correlation) networks were used as brief case studies. 

Using (geo)graphs, as a concept, and G4G and GeoCNet, as tools, you are able to insert and manipulate graphs into GIS, the most appropriate computational environment to handle geographical data. Then, you are able to easily representing graphs and their properties on maps, composing with different layers and allowing several spatial analysis.

Among the perspectives of this research are the improvement of the web interface for the tool, and more detailed case studies, producing thematic maps for the nodes, edges and properties (topological measurements).


\bibliography{sample}


\section*{Acknowledgements}

Research was partially supported by grant 454267/2014-2 of the Brazilian National Council for Scientific and Technological Development (CNPq) and by the grant 2015/50122-0 São Paulo Research Foundation (FAPESP) and DFG-IRTG 1740/2. C.N.S was partially supported by a PDJ Postdoctoral grant 23038.000776/2017-54 by the Coordination for the Improvement of Higher Education Personnel (CAPES). The authors sincerely thank Dr. Marcos Quiles, Dr. Elbert Macau and Dr. Tristan Pryer for their kind comments during this study.


\section*{Author contributions statement}

L.B.L. conceived the concept of (geo)graphs, A.A.S.J., M.R. and J.D.S implemented the main computational tools, A.A.S.J., M.R., J.D.S, O.A.C. and W.S. conducted the experiments, L.B.L. and C.N.S. analyzed the results. All author reviewed the manuscript. 





\end{document}